\documentclass[aps,10pt,nofootinbib,preprintnumbers,showpacs]{revtex4}

\usepackage{amssymb, amsfonts, amsmath, bm}
\usepackage[
]{graphicx,color}
\usepackage[
]{hyperref}
\hypersetup{
 bookmarksnumbered=true,%
 bookmarksopen=true,%
 colorlinks=true,%
 linkcolor=blue,%
 citecolor=red,
 }

\begin{document}
\title{{\Large Scale invariance and constants of motion}}
\author{{\large Takahisa Igata}}
\email{igata@rikkyo.ac.jp}
\affiliation{Department of Physics, Rikkyo University, Toshima, Tokyo 171-8501, Japan}
\date{\today}
\preprint{RUP-18-12}

\begin{abstract}
Scale invariance in the theory of 
classical mechanics can be induced from 
the scale invariance of background fields. 
In this paper we consider the relation between the 
scale invariance and the constants of particle motion 
in a self-similar spacetime, only in which 
the symmetry is well-defined and is generated by a homothetic vector. 
Relaxing the usual conservation condition 
by the Hamiltonian constraint in a particle system, 
we obtain a conservation law 
holding only on the constraint surface in the phase space. 
By the conservation law, 
we characterize constants of motion 
associated with the scale invariance 
not only for massless particles but for 
massive particles 
and classify the condition for the existence of the
constants of motion. 
Furthermore, we find the explicit form of 
the constants of motion by solving the conservation equations. 
\end{abstract}

\maketitle

\section{Introduction}
\label{sec:1}
The symmetry of external fields can induce 
the symmetry of a theory on the background, 
which is embodied as that of the action integral. 
As is well known, 
the action invariance under a symmetric continuous transformation 
leads to a conservation law via Noether's theorem 
(see, e.g., Ref.~\cite{Banados:2016zim}). 
Since the conservation law is based on symmetry, 
it has a universal role in physics. 
Furthermore, symmetry can also define symmetric configurations of test fields 
and extended objects~\cite{Igata:2009dr, Igata:2009fd, Igata:2012kx, Kozaki:2014aaa, Kinoshita:2016lqd, Morisawa:2017lpj, Kinoshita:2017mio}.

In classical particle mechanics with continuous symmetry, 
a particle has a conserved quantity throughout the motion 
associated with the symmetry---a constant of motion. 
In particular, geometrical symmetry in a spacetime can be the 
origin of symmetry in a particle system. 
If a spacetime metric admits an isometry, 
then the generator (i.e., the Killing vector) 
yields a constant of motion. 
This symmetry is generalized to 
spacetime hidden symmetries such as 
higher-rank Killing--Stackel tensors or 
Killing--Yano forms~(see, e.g., Refs.~\cite{Yasui:2011pr, Frolov:2017kze}). 
In addition, if a metric admits 
a conformal isometry, then 
it induces the conformal invariance on the action of a massless particle. 
This means that 
a massless particle can have a constant of motion 
associated with a conformal Killing vector. 
Note that, however, 
the conformal Killing vector is not related to 
the constant of motion of a massive particle in general. 
Even if a particle is subject to external forces, 
it can have a constant of motion as long as the external 
fields share the spacetime symmetry~\cite{Sommers:1973, Gibbons:1993ap, vanHolten:2006xq, Ngome:2009pa, Igata:2010ny, Visinescu:2011if, Cariglia:2014dfa, Cariglia:2014dwa}.

The scale invariance of a metric is a special class of the conformal symmetry
and is known as the spacetime self-similarity. 
In the context of relativity, the self-similar spacetime 
appears as a critical point of the critical phenomena in 
gravitational collapse~\cite{Choptuik:1992jv, Koike:1995jm} 
or describes the asymptotic behavior of spatially homogeneous models in cosmology and 
spherically symmetric models~(see, e.g., Ref.~\cite{Carr:1998at} and references therein). 
More recently, the notion of self-similar strings was proposed 
in the classical string theory~\cite{Igata:2016uvp}. 
The spacetime self-similarity induces the scale invariance of a theory on the background, i.e., 
the invariance of the action under the scale transformation of variables. 
However, the corresponding constants of motion for a massive particle 
are less well known~\cite{Papadopoulos:2000ka} because, 
as mentioned above, the constant of motion associated with 
the conformal invariance exists only for massless particles. 
On the other hand, in nonrelativistic mechanics, the constant of motion associated with scale invariance exists, e.g., 
in the conformal particle system~\cite{deAlfaro:1976vlx} 
or in the Kepler problem of 5D Newtonian gravity, etc. (see Appendix~\ref{sec:A}). 
The main issue of this paper is such inconsistency.

The way to solve this inconsistency is to 
consider that 
a particle system in gravitational theories 
is a constraint system. 
The constraint arises from 
the reparameterization invariance of the action. 
We can utilize it to 
relax the conservation condition for a 
constant of motion, i.e., 
we restrict the conservation condition
holding only on the constraint surface in the phase space. 
The purpose of this paper is to characterize the 
constant of massive particle motion associated with scaling symmetry 
as the quantity conserved only on the constraint surface.

This paper is organized as follows. 
In the following section, we reformulate the conservation law 
for a dynamical quantity in the Hamiltonian formalism 
by taking into account the constraint condition. 
In Sect.~\ref{sec:3}, 
assuming that the spacetime metric 
and the other fields are scale invariant, 
we discuss the existence of a constant of motion associated with 
the scaling symmetry of the theory 
and solve the conservation equation. 
In Sect.~\ref{sec:4}, we summarize our results. 
Throughout this paper, 
we use geometrized units, in which $G=1$ and $c=1$, and 
an abstract index notation for tensors but 
boldface letters without indices to designate differential forms~\cite{Wald:1984rg}.

We summarize our notation in what follows. 
We denote the symbol $v_n$ as the type-$(n, 0)$ symmetric tensor $v^{a_1\cdots a_n}=v^{(a_1\cdots a_n)}$, where the parenthesis of the indices means totally symmetrization, 
$v_1$ is a vector, and $v_0$ is a scalar in particular. 
We denote the contraction of $v_n$ with $n$ pieces of a dual vector $p_a$ as
\begin{align}
v_n\cdot p^n=v^{a_1a_2\cdots a_n} p_{a_1} p_{a_2}\cdots p_{a_n}.
\end{align}
If we identify $p_a$ with a canonical momentum in a particle system, 
then $V=v_n\cdot p^n$ and $W=w_m \cdot p^m$ are 
dynamical quantities. 
We define the Poisson bracket of $V$ and $W$ as 
\begin{align}
\{V, W\}=\frac{\partial V}{\partial p_\mu}\frac{\partial W}{\partial x^\mu}-\frac{\partial V}{\partial x^\mu}\frac{\partial W}{\partial p_\mu}
=\left[\:\!v_n, w_m\:\!\right]\cdot p^{n+m-1},
\end{align}
where we have introduced the 
symmetric Schouten--Nijenhuis bracket of
the symmetric tensors $v_n$ and $w_m$ defined as 
\begin{align}
\left[\:\!v_n, w_m \:\!\right]^{a_1 \cdots a_{n-1}b_1 \cdots b_{m-1}c}
&= n\:\! v^{\left(a_1 \cdots a_{n-1} |\:\!d\:\!|\right.}
\nabla_d 
w^{\left.b_1 \cdots b_{m-1}c\right)}-
m \:\!w^{\left(b_1 \cdots b_{m-1} |\:\!d\:\!|\right.}\nabla_d v^{\left.
a_1 \cdots a_{n-1}c\right)}.
\end{align}
Note that this equation holds for any derivative operator. 
In the case where $n=1$ and $m=1$, 
this reduces to the Lie bracket of the vectors $v_1$ and $w_1$. 
We denote the exterior derivative of a form as $\mathrm{d}$. 
The dot product of a 1-form, $\bm{\xi}=\xi_a$, 
and a $p$-form, $\bm{F}=F_{[\:\!a_1\cdots a_p\:\!]}$, 
denotes $\bm{\xi}\cdot \bm{F}=\xi^{a_1}F_{[\:\!a_1\cdots a_p\:\!]}$.

\section{Formulation}
\label{sec:2}

We consider classical particle mechanics 
in a $D$-dimensional spacetime $(M, g_{ab})$. 
Let $H$ be the Hamiltonian of a particle. 
In an appropriate form of $H$, 
the particle must satisfy the Hamiltonian constraint $H=0$ 
because of the reparameterization invariance of the world line. 
The Hamiltonian should vanish only if the 
constraint holds, i.e., 
$H$ is \textit{weakly zero}, which is expressed as
\begin{align}
H\approx 0.  
\end{align}
We should apply the weak equality 
after evaluating derivatives or Poisson brackets.

In the particle system, we consider a dynamical quantity $C$, 
which depends on local coordinates $x^\mu$ in $M$, 
a canonical momentum $p_\mu$ conjugate to $x^\mu$, 
and a parameter $\tau$ on the world line~$\gamma$, 
so that $C=C(x, p; \tau)$. 
We assume that 
$C$ is a constant of motion, i.e., 
the time evolution of $C$ via the Hamilton equations vanishes. 
Then, $C$ must satisfy 
\begin{align}
\label{eq:conservation_law}
\dot{C}\approx 0,  
\end{align}
where the dot denotes the derivative with respect to $\tau$. 
Note that this condition is weaker than that 
required in the total phase space. 
Namely, we have relaxed the condition $\dot{C}=0$ 
by using the constraint condition because 
it is sufficient for a constant of motion to be conserved 
only on the constraint surface $H=0$. 
We can express the condition \eqref{eq:conservation_law} 
by the equation holding in the total phase space as 
\begin{align}
\label{eq:conservation}
\{H, C\}+\frac{\partial C}{\partial \tau}=\lambda \:\!H,
\end{align}
where $\lambda=\lambda(x, p, \tau)$.

Now, we assume that $H$ is a second-degree polynomial 
in the canonical momentum $p_a$ as
\begin{align}
\label{eq:Hh}
H=h_2 \cdot p^2+h_1\cdot p^1+h_0. 
\end{align}
and $C$ is a 
first-degree polynomial in $p_a$ as
\begin{align}
C=c_1\cdot p^1+c_0,
\end{align}
where $c_1$ and $c_0$ 
depend on both $x^\mu$ and $\tau$. 
Substituting these into Eq.~\eqref{eq:conservation} 
with the polynomial expansion of $\lambda$ in $p_a$, 
\begin{align}
\lambda=\sum_{n} \lambda_n \cdot p^n,
\end{align} 
we obtain the following equations by setting 
the coefficients of its momentum-expansion equal to zero as
\begin{align}
\label{eq:KH1}
&[\:\!h_2, c_1\:\!]
=\lambda_0 h_2,
\\[1.3mm]
\label{eq:KH2}
&[\:\!h_2, c_0\:\!]+[\:\!h_1, c_1\:\!]
+\frac{\partial c_1}{\partial \tau}
=\lambda_0 h_1,
\\
\label{eq:KH3}
&[\:\!h_1, c_0\:\!]+[\:\!h_0, c_1\:\!]+\frac{\partial c_0}{\partial \tau}
=\lambda_0 h_0. 
\end{align}
We find that $\lambda=\lambda_0$ 
because $\{\lambda_n\}_{n=1, 2, \ldots}$ vanish thanks to 
the equations higher than rank 3. 
These equations 
are a generalization of the Killing hierarchy~\cite{Igata:2010ny}.   
Equations~\eqref{eq:KH1} and \eqref{eq:KH2} are 
the eqautions of $c_1$ and $c_0$, respectively, while 
Eq.~\eqref{eq:KH3} implies the 
consistency condition for the existence of $c_1$ and $c_0$. 
It is worth pointing out that 
we can generalize these hierarchical equations further to 
the higher-degree polynomial of $C$.

\section{Constants of motion in a self-similar spacetime}
\label{sec:3}
Let $(M, g_{ab})$ be a self-similar spacetime, i.e., 
the metric $g_{ab}$ admits a homothetic vector $\xi^a$, defined by%
\footnote{See, e.g., Ref.~\cite{Yano1955} 
for the geometrical properties of the homothetic vector.  }
\begin{align}
\label{eq:HVeq}
\pounds_\xi g_{ab}=2\:\!g_{ab}, 
\end{align}
where $\pounds_\xi$ is the Lie derivative with respect to $\xi^a$. 
The homothetic vector is unique 
up to the addition of a Killing vector~\cite{Eardley:1973fm}. 
In such a spacetime, 
we focus on a particle that is subject to external forces from 
a $1$-form gauge field~$\bm{A}$ and a scalar potential~$V \geq 0$. 
Let $\bm{F}$ be the field strength 2-form of $\bm{A}$, i.e., 
$\bm{F}=\mathrm{d}\bm{A}$. 
We assume that 
$\bm{F}$ and $V$ 
have self-similarity 
induced by the spacetime self-similarity as%
\footnote{
Since a gauge-invariant quantity is not $\bm{A}$ but $\bm{F}$, 
it is reasonable that the condition of self-similarity is written in terms of $\bm{F}$. 
Even if $\bm{A}$ is not a gauge field, 
the self-similarity condition for $\bm{A}$ 
\begin{align}
\pounds_\xi \bm{A}=\alpha \bm{A}
\end{align}
leads to Eq.~\eqref{eq:SSF}, 
where we have used the fact that $\mathrm{d}$ commutes with $\pounds_\xi$.
Therefore, the following discussions hold in the same way. }
\begin{align}
\label{eq:SSF}
&\pounds_\xi \bm{F}=\alpha\:\! \bm{F},
\\
\label{eq:SSV}
&\pounds_\xi V=\beta \:\!V, 
\end{align}
where $\alpha$ and $\beta$ are constants. 
The scale invariance of background fields is based on 
Eqs.~\eqref{eq:HVeq}, \eqref{eq:SSF}, and \eqref{eq:SSV}. 
According to Ref.~\cite{Gralla:2017lto}, we refer to 
$\bm{F}$ as a self-similar 2-form with weight $\alpha$ and 
$V$ as a self-similar scalar with weight $\beta$. 
If $\bm{F}=0$, then $\alpha$ is indefinite, 
and if $V=0$,  then $\beta$ is indefinite. 
We consider the particle system governed by the Polyakov action 
\begin{align}
\label{eq:S}
S=\int \mathrm{d}\tau \left[\:\!
\frac{1}{2N} \:\!g_{ab} u^a u^b
+A_a u^a
-N V
\:\!\right], 
\end{align}
where the components of $u^a$ are $u^\mu=\dot{x}^\mu$ 
and $N>0$ is an auxiliary variable. 
Note that $S$ is invariant under the gauge transformation 
$\tau \to \tau'$ and $N \to N' =(\mathrm{d}\tau/\mathrm{d}\tau') N$, 
i.e., $S$ is reparameterization invariant. 
The action principle yields the equations of motion 
\begin{align}
\label{eq:eom}
&u^b \nabla_b u^a
=N F^a{}_{b} u^b
-N^2 g^{ab}\nabla_b V
+(\ln N)^{\dot{}} \:\!u^a. 
\end{align}
In addition, the variation of $S$ with respect to $N$ leads to the 
constraint equation%
\footnote{
One can deparameterize the theory by solving the constraint for $N$, and then 
$S$ is of the form
\begin{align}
\label{eq:deparaS}
S
=\int_\gamma \left[\:\!
-\sqrt{2\:\!V\:\! g_{\mu\nu} \mathrm{d}x^\mu \:\!\mathrm{d}x^\nu
}+A_\mu \:\!\mathrm{d}x^\mu
\:\!\right].
\end{align}
Hence, we can regard the particle system in Eq.~\eqref{eq:S}
as the system in a spacetime $(M, 2Vg_{ab})$ without scalar potential forces. 
}
\begin{align}
\label{eq:constraint}
&g_{ab} u^a u^b =-2N^2 V. 
\end{align}
This equation implies that 
$N$ is relevant to the 1D induced metric on $\gamma$. 
If $V=0$, the tangent $u^a$ becomes null, i.e., 
the particle is massless. 
The Hamiltonian of this system is of the form
\begin{align}
\label{eq:H}
&H=N\left[\,
\frac{1}{2}\:\!
g^{ab} \left(p_a -A_a \right)\left(p_b-A_b\right)+V
\:\!\right],
\end{align}
where the components $ p_\mu$ of $p_a=N^{-1} g_{ab}u^b+A_a$ are the canonical momentum 
conjugate to $x^\mu$.
The constraint~\eqref{eq:constraint} leads to 
$H=0$ in the Hamiltonian formalism. 
Without loss of generality, 
we assume that $N=1$ in what follows. 

\medskip

In the following, we solve 
the hierarchical equations~\eqref{eq:KH1}--\eqref{eq:KH3}
and obtain a nontrivial solution relevant to the homothetic vector $\xi^a$. 
With the Hamiltonian~\eqref{eq:H}, 
we can reduce 
Eq.~\eqref{eq:KH1} to 
the conformal Killing equation for $c_1^a$ and 
solve it by using $\xi^a$ in the form
\begin{align}
\label{eq:c1}
c_1^a=f(\tau)\:\!\xi^a,
\quad 
\lambda_0=2\:\!f(\tau),
\end{align}
where $f$ is a positive function of $\tau$ and 
we can arbitrarily choose the scale of $f(\tau)$. 
With these forms, Eq.~\eqref{eq:KH2} becomes 
\begin{align}
\label{eq:dc0}
\mathrm{d}c_0=
-f(\tau)\left[\:\!
\bm{\xi}\cdot \bm{F}+\mathrm{d}(\bm{A}\cdot \bm{\xi})
+\frac{\dot{f}(\tau)}{f(\tau)}\:\!\bm{\xi}\:\!\right]. 
\end{align}
The integrability condition for $c_0$, i.e.,  
$\mathrm{d}^2c_0=0$, leads to 
\begin{align}
\label{eq:ddc0}
\mathrm{d}\left[\:\!
\bm{\xi}\cdot \bm{F}+\frac{\dot{f}(\tau)}{f(\tau)}
\bm{\xi}
\:\!\right]=0.
\end{align}
We can replace this first term by using 
Cartan's identity%
\footnote{
This is the relation between the Lie derivative $\pounds_{\xi}$ 
with respect to $ \xi^a$ 
and the exterior derivative $\mathrm{d}$
in differential forms: 
\begin{align}
\pounds_\xi \bm{F}
=\bm{\xi}\cdot \mathrm{d}\bm{F}
+\mathrm{d} (\bm{\xi}\cdot \bm{F}),
\end{align}
where $\bm{F}$ is an arbitrary differential form. 
When $\bm{F}$ is the field strength, the first term on the right-hand side vanishes because $\mathrm{d}\bm{F}=0$. 
} 
and the self-similarity~\eqref{eq:SSF} as
\begin{align}
\label{eq:integrability}
\alpha\:\!\bm{F}
+\frac{\dot{f}(\tau)}{f(\tau)}
\mathrm{d}\:\!\bm{\xi}=0. 
\end{align}
In the following subsections, 
we discuss each of the following cases: 
(A) $\alpha=0$ and $f(\tau)=1$, (B) $\alpha=0$ and 
$ \mathrm{d}\:\!\bm{\xi}=0$,
and (C) $\alpha\neq 0$. 
Even if we set $\bm{F}=0$ in Cases~(A) and (B), 
where $\alpha$ is no longer definite, 
all the results will remain unchanged.

\subsection{$\alpha=0$ and $f(\tau)=1$}
\label{sec:3A}
Then, there locally exists a solution to Eq.~\eqref{eq:dc0} in the form
\begin{align}
\label{eq:c0in3A}
c_0=
\varphi(x)%
-\bm{\xi}\cdot \bm{A}
+l(\tau),
\end{align}
where $l$ is an arbitrary function of $\tau$, 
and $\varphi$ is a potential function of $x^\mu$ that satisfies
\begin{align}
\bm{\xi}\cdot \bm{F}=-\mathrm{d}\varphi
\end{align}
due to Eq.~\eqref{eq:ddc0}.
In addition, $c_1$, $c_0$, and $\lambda_0$ 
in Eqs.~\eqref{eq:c1} and \eqref{eq:c0in3A}
must satisfy the consistency condition~\eqref{eq:KH3}, 
which reduces as
\begin{align}
\label{eq:Vin3A}
\left(\beta +2\right)V=\kappa, 
\quad 
l(\tau)=\kappa \:\!\tau, 
\end{align}
where we have used Eq.~\eqref{eq:SSV}, and 
$\kappa$ is the separation constant of variables. 
The Lie derivative of the first with respect to $\xi^a$ 
and Eq.~\eqref{eq:SSV} yield
\begin{align}
\label{eq:betaVcondition}
\beta\left(\beta+2\right)V=0. 
\end{align} 
Hence, if any one of the following conditions holds: 
\begin{itemize}
\item[(i)] $V=\kappa/2>0$ and $\beta=0$,
\item[(ii)] $V>0$, $\beta=-2$, and $\kappa=0$,
\item[(iii)] $V=0$ and $ \kappa=0$,
\end{itemize}
then we obtain the constant of motion 
associated with the scale invariance in the form
\begin{align}
\label{eq:Cin3A}
C=\bm{u}\cdot \bm{\xi}
+\varphi(x)+\kappa\:\!\tau.
\end{align}
In Case~(i), the particle is massive 
and is not subject to the scalar potential force, 
so that $V$ is necessarily a self-similar potential field with weight $0$. 
Even for a massive particle, 
we obtain the constant of motion associated with the scale invariance 
thanks to the presence of the last term in Eq.~\eqref{eq:Cin3A}, 
which is one of the constants of motion in question in Sect.~\ref{sec:1}. 
Indeed, this type of constant 
with $\bm{F}=0$ was utilized 
to integrate the equations of motion of a freely falling massive particle system (i.e., a geodesic system) in 
a self-similar 4D Vaidya spacetime~\cite{Joshi:1992vr}. 
In Case~(ii), 
the particle is massive and can be subject to the nontrivial potential force, 
and $V$ is a self-similar potential field with weight $(-2)$. 
We can understand the specialty of $\beta=-2$ as follows. 
When we regard the particle system~\eqref{eq:S} as 
the system in a spacetime $(M, 2Vg_{ab})$ as seen in Eq.~\eqref{eq:deparaS}, 
we find that the constant of motion is not associated with 
the homothetic vector but the Killing vector in $(M, 2Vg_{ab})$ because 
\begin{align}
\pounds_\xi \left(V g_{ab}\right)=\left(\beta+2\right) Vg_{ab}. 
\end{align}
Therefore, from the point of view in $(M, 2Vg_{ab})$, 
$C$ reduces to the familiar one. 
In Case~(iii), 
the particle is massless, so that $\beta$ is indefinite. 
Then $C$ is a familiar constant of motion 
associated with the conformal Killing vector.

\subsection{$\alpha=0$ and $\mathrm{d}\:\!\mbox{\boldmath $\xi$}=0$ }
\label{sec:3B}
To seek constants of motion in addition to those in Sect.~\ref{sec:3A}, 
and also to find constants of motion for other values of $\beta$ rather than 
that in Sect.~\ref{sec:3A}, 
we solve the remaining equations in the case where 
$\bm{F}$ is a self-similar 2-form with weight $0$ (i.e., $\alpha=0$), 
and $\bm{\xi}$ is closed (i.e., $\mathrm{d}\:\!\bm{\xi}=0$). 
The latter condition implies 
the local existence of the potential function $\zeta$ of $\bm{\xi}$, i.e., 
\begin{align}
\bm{\xi}=\mathrm{d} \zeta,
\end{align}
where $\zeta$ is not constant.
On account of the integrability, there locally exists a solution in the form
\begin{align}
\label{eq:c0in3B}
c_0=f(\tau)\left(\varphi(x)-\bm{\xi}\cdot \bm{A}\right)
-\dot{f}(\tau) \:\!\zeta(x) 
+l(\tau),
\end{align}
where $l$ is an arbitrary function of $\tau$ and $\varphi$ is 
a potential function that satisfies 
\begin{align}
\label{eq:phi}
\bm{\xi}\cdot \bm{F}=-\mathrm{d}\varphi. 
\end{align}
The quantities $c_1$, $c_0$, and $\lambda_0$ 
in Eqs.~\eqref{eq:c1} and \eqref{eq:c0in3B}
must satisfy the consistency condition~\eqref{eq:KH3}, 
which reduces to
\begin{align}
\label{eq:Vin3B}
\left(\beta+2\right)V
=\frac{1}{f(\tau)}\left[\:\!
\dot{f}(\tau) \:\!\varphi
-\ddot{f}(\tau)\:\!\zeta
+\dot{l}(\tau)
\:\!\right], 
\end{align}
where we have used Eq.~\eqref{eq:SSV}. 
It is useful to obtain the gradient of this equation 
\begin{align}
\label{eq:gradV}
\left(\beta+2\right)\mathrm{d}V
=\frac{\dot{f}(\tau)}{f(\tau)}\:\!\mathrm{d}\varphi
-\frac{\ddot{f}(\tau)}{f(\tau)}\:\!\mathrm{d}\zeta
\end{align}
and the Lie derivative of Eq.~\eqref{eq:Vin3B} with respect to $\xi^a$ 
\begin{align}
\label{eq:betaVin3B}
\beta\left(\beta+2\right)V
=-\frac{\ddot{f}(\tau)}{f(\tau)} \bm{\xi}\cdot \bm{\xi},
\end{align}
where $\pounds_\xi \varphi=0$ due to Eq.~\eqref{eq:phi}.
Finding a nontrivial pair of $f(\tau)$ and $l(\tau)$ 
that satisfy Eq.~\eqref{eq:Vin3B}, 
we obtain a constant of motion associated with 
the scale invariance in the form
\begin{align}
\label{eq:Cin3B}
C=f(\tau)\:\!\left[\:\!
\bm{u}\cdot \bm{\xi}
+\varphi(x)\:\!\right] -\dot{f}(\tau) \:\!\zeta(x)+l(\tau). 
\end{align}
We ignore a constant term of $f(\tau)$ in this subsection 
because such contribution in $C$ coincides with 
Eq.~\eqref{eq:Cin3A} in Sect.~\ref{sec:3A}. 
Now, classifying cases according to 
whether $\beta\left(\beta+2\right)V$ is zero or not, 
we analyze the reduced consistency condition~\eqref{eq:Vin3B} in detail 
and obtain the explicit form of Eq.~\eqref{eq:Cin3B} and 
necessary conditions for its existence.

\subsubsection{$\beta\left(\beta+2\right)V=0$}
This condition coincides with the necessary condition~\eqref{eq:betaVcondition} in Sect.~\ref{sec:3A}. 
From Eq.~\eqref{eq:betaVin3B}, 
we have the two possibilities: 
$\ddot{f}(\tau)=0$ or $\bm{\xi}\cdot \bm{\xi}=0$, 
and analyze these in turn.

We focus on $\ddot{f}(\tau)=0$, which is integrated as 
\begin{align}
f(\tau)=\tau,
\end{align}
where, without loss of generality, we have fixed the scale of $f(\tau)$. 
Furthermore, from Eqs.~\eqref{eq:Vin3B} and \eqref{eq:gradV}, 
we find that 
\begin{align}
\varphi=0,
\quad
\kappa=\left(\beta+2\right)V, 
\quad 
l(\tau)=\frac{\kappa}{2}\tau^2,
\end{align}
where $\kappa$ is a separation constant and, 
without loss of generality, 
we choose the constant term of $\varphi$ to be zero. 
The second equation coincides with the first in Eq.~\eqref{eq:Vin3A}, 
and hence the same conditions in Cases~(i)--(iii) in Sect.~\ref{sec:3A} hold. 
Finally, the constant of motion $C$ in Eq.~\eqref{eq:Cin3B} reduces to 
\begin{align}
\label{eq:Cin3B-1}
C
=\tau\:\! \bm{u}\cdot \bm{\xi}
-\zeta(x)
+\frac{\kappa}{2}\tau^2. 
\end{align}
Hence, if the conditions $\alpha=0$, 
$\mathrm{d}\bm{\xi}=0$, $\varphi=0$, 
and any one of the conditions (i)--(iii) in Sect.~\ref{sec:3A} are satisfied, 
then we obtain 
the two constants of motion~\eqref{eq:Cin3A} and 
\eqref{eq:Cin3B-1}. 
These results hold regardless of whether $\xi^a$ is null or not. 
In Cases~(i) and (ii), $C$ is the constant of motion
for a massive particle. 
Even for a massless particle in Case~(iii), 
$C$ is still nontrivial because of the dependence of $\tau$.

Next, we focus on the remaining case where 
$\ddot{f}(\tau)\neq 0$ and $\bm{\xi\cdot\xi}=0$. 
From Eq.~\eqref{eq:gradV} and 
its derivative with respect to $\tau$, we have
\begin{align}
\varphi=s\:\!\zeta, 
\quad
\ddot{f}(\tau)=s\:\!\dot{f}(\tau)+a\left(a-s\right)f(\tau),
\quad
\left(\beta+2\right) \mathrm{d}V=
-a\left(a-s\right)\mathrm{d}\zeta, 
\end{align}
where $s$ is a separation constant and  
$a$ and $\kappa$ are integration constants. 
We assume that $a$ is not zero but can take a complex value. 
In the case $\beta=-2$ or $V=0$, we find $a=s$. 
Then we can solve the second equation by
\begin{align}
\label{eq:f}
f(\tau)=e^{a \tau}. 
\end{align}
Substituting these results into Eq.~\eqref{eq:Vin3B}, we have
\begin{align}
\left(\beta+2\right) V
=\kappa-a\left(a-s\right)\zeta,
\quad 
l(\tau)=\frac{\kappa}{a}e^{a\tau}. 
\end{align}
where $\kappa$ is a separation constant. 
Finally we obtain the explicit form of Eq.~\eqref{eq:Cin3B} as
\begin{align}
\label{eq:Cin3B-1null}
C=e^{a\tau}\left[\:\!
\bm{u}\cdot \bm{\xi}+\varphi(x)-a\:\!\zeta(x)+\frac{\kappa}{a}
\:\!\right].
\end{align}
Note that this is not compatible with Eq.~\eqref{eq:Cin3B-1}. 
In the case $\beta=0$ and $a\neq s$, 
though there are two constants of motion corresponding to 
two linearly independent solutions~\eqref{eq:f} with different values of $a$, 
we find that 
these coincide with each other by rescaling $\tau$. 
If the conditions $\alpha=0$, $\mathrm{d}\bm{\xi}=0$, $\bm{\xi}\cdot \bm{\xi}=0$, $\varphi=s\:\!\zeta$, and any one of the conditions (i)--(iii) in Sect.~\ref{sec:3A} are satisfied, 
then we obtain the two constants of motion~\eqref{eq:Cin3A} and \eqref{eq:Cin3B-1null}.

\subsubsection{$\beta\left(\beta+2\right)V\neq0$}
In this case we obtain the separated equations 
from Eq.~\eqref{eq:betaVin3B} as%
\footnote{The potential proportional to the norm of $\xi^a$ 
appears in the context of self-similar strings~\cite{Igata:2016uvp}.}
\begin{align}
\label{eq:betaVin3B-2}
\beta \left(\beta+2\right) V=-s^2\:\!\bm{\xi}\cdot \bm{\xi},
\quad 
f(\tau)=e^{s\:\! \tau}, 
\end{align}
where $s^2$ is a nonzero real constant. 
The norm $\bm{\xi}\cdot \bm{\xi}$ should not be constant 
because the resulting condition 
$\beta=0$ is contrary to the assumption here. 
Hence, we find the relations of each potential 
from Eqs.~\eqref{eq:Vin3B} and \eqref{eq:betaVin3B-2} as
\begin{align}
\left(\beta+2\right)V
=\kappa+s\:\!\varphi-s^2\:\!\zeta,
\quad l(\tau)
=\frac{\kappa}{s}e^{s\:\! \tau}. 
\end{align}
With these constraints for $V$, $\varphi$, and $\zeta$, we obtain 
the constant of motion $C$ in Eq.~\eqref{eq:Cin3B} as
\begin{align}
C=e^{s\:\!\tau}\left[\:\!
\bm{u}\cdot \bm{\xi}
+\varphi(x)-s\:\!\zeta(x)
+\frac{\kappa}{s}
\:\!\right]. 
\end{align}
Though we obtain two constants of motion corresponding to 
two linearly independent solutions, 
we find that these coincide with each other by rescaling the parameter $\tau$.

\subsection{$\alpha\neq 0$}
\label{sec:3C}
We solve the remaining equations in the case 
where $\bm{F}$ is a self-similar 2-form with 
nonzero weight $\alpha$. 
We obtain the following separated equation from the 
integrability condition~\eqref{eq:integrability}:
\begin{align}
\label{eq:integrabilityin3C}
\alpha\:\!\bm{F}+s\:\!\mathrm{d}\:\!\bm{\xi}=0,
\quad
f(\tau)=e^{s\tau},
\end{align}
where $s$ is a nonzero separation constant. 
Note that $ \bm{F}$ is nonzero and 
$\bm{\xi}$ should not be closed. 
From the first in Eq.~\eqref{eq:integrabilityin3C}, 
up to the gauge transformation, 
$\bm{A}$ must be proportional to $\bm{\xi}$:
\begin{align}
\bm{A}=-\frac{s}{\alpha} \bm{\xi}. 
\end{align}
On account of the integrability, 
there locally exists a solution in the form
\begin{align}
\label{eq:c0an0}
c_0=e^{s\tau}
\left[\:\!\varphi(x)-\bm{\xi}\cdot \bm{A}
\:\!\right]+l(\tau),
\end{align}
where $l$ is an arbitrary function of $\tau$ and $\varphi$ is a potential function that  satisfies
\begin{align}
\label{eq:phi3C}
\bm{\xi}\cdot \bm{F}+s\:\!\bm{\xi}=-\mathrm{d}\varphi. 
\end{align}
The quantities $c_1$, $c_0$, and $\lambda_0$ in Eqs.~\eqref{eq:c1} and \eqref{eq:c0an0}
must satisfy the consistency condition~\eqref{eq:KH3}, 
which reduces to
\begin{align}
\left(\beta+2\right)V=s\:\!\varphi,\quad
l(\tau)=0.
\end{align}
The Lie derivative of the first equation with respect to $\xi^a$ yields 
\begin{align}
\beta\left(\beta+2\right) V=-s^2\:\!\bm{\xi}\cdot \bm{\xi},
\end{align}
where we have used Eqs.~\eqref{eq:SSV} and \eqref{eq:phi3C}. 
Finally, we obtain a constant of motion associated with 
the scale invariance in the form
\begin{align}
C=e^{s\tau}\left[\:\!
\bm{u}\cdot \bm{\xi}
+\varphi(x)
\:\!\right]
\end{align}
where $\varphi$ is given by the integration of Eq.~\eqref{eq:phi3C}.

\section{Summary}
\label{sec:4}

We have considered constants of particle motion 
associated with scale invariance in classical mechanics 
on a curved spacetime. 
The scale invariance can be induced from the 
scale symmetry of background fields. 
If both the mechanics and the background fields share 
continuous scale symmetry, 
a particle can have a constant of motion associated with it. 
On the other hand, 
a conformal Killing vector generating conformal symmetry 
is related to a constant of massless particle motion in general. 
Since a homothetic vector generating scale symmetry 
is classified in the class of conformal symmetry, 
it was unclear how to characterize a 
constant of massive particle motion 
associated with a homothetic vector. 
In this paper, 
we have made it clear that,  
even for massive particles, 
a constant of motion associated with the homothetic vector 
exists by virtue of the reparameterization invariance of a particle system.  
We have utilized the constraint condition relevant to 
the reparameterization invariance 
to relax the conservation condition for constants of motion. 
The conservation law that we have obtained takes the form of 
hierarchical equations. 
Solving these equations to particle mechanics in background fields with 
scale invariance, 
we have obtained constants of motion associated with the scale invariance 
and have classified the conditions for the existence 
by the self-similar weight of the background fields. 
Consequently, 
we have found that 
constants of massive particle motion associated with 
the scale invariance must depend explicitly on 
a parameter on the world line.

\begin{acknowledgments}
The author thanks 
T.~Harada, M.~Nagashima, and T.~Tanaka for useful comments. 
This work was supported by the MEXT-Supported Program for the Strategic Research Foundation at Private Universities, 2014--2017 (S1411024). 
\end{acknowledgments}

\appendix

\section{Scale invariance in nonrelativistic mechanics}
\label{sec:A}
We briefly review nonrelativistic classical mechanics with scale invariance. 
Let $\bm{r}(t)$ be the position vector at time $t$ of a particle with mass $m$ 
and let $S$ be the action 
\begin{align}
S=\int \mathrm{d} t\:\!\bigg[\:\!
\frac{m}{2}
\left|\:\!
\dot{\bm{r}}(t)
\:\!\right|^2
-V(\bm{r}(t))
\:\!\bigg],
\end{align}
where $\dot{\bm{r}}=\mathrm{d}\bm{r}/\mathrm{d} t$, and 
$V(\bm{r})$ is a potential function. 
We assume that $V(\bm{r})$ is 
a homogeneous function of degree $k$, i.e., 
\begin{align}
V(\alpha\:\!\bm{r})=\alpha^k\:\!V(\bm{r}),
\end{align}
where $\alpha$ is a constant. 
Then the scale transformation 
\begin{align}
\bm{r}\to \tilde{\bm{r}}=\alpha \:\!\bm{r},
\quad
t\to \tilde{t}=\alpha^{1-k/2} t
\end{align}
to the action yields 
the transformed action $\tilde{S}$ that is related to $S$ as 
\begin{align}
\label{eq:tildeS}
\tilde{S}=\alpha^{1+k/2}\:\!S.
\end{align}
In general, Eq.~\eqref{eq:tildeS} only implies the scale invariance of the Lagrangian.  
In the Kepler problem of 4D Newtonian gravity, i.e., $V\propto |\:\!\bm{r}\:\!|^{-1}$, 
this invariance leads to Kepler's 3rd law (see, e.g., Ref.~\cite{Landau:1976}). 
When $k=-2$, however, 
the action itself is scale invariant. 
As an example of $V$ with $k=-2$, 
the potential $V\propto |\:\!\bm{r}\:\!|^{-2}$
is found in a conformal particle system or 
necessarily appears in the Kepler problem of 5D Newtonian gravity. 
In such case,  
we obtain a constant of motion 
associated with the scale invariance from Noether's theorem as
\begin{align}
C=\mbox{\boldmath $p$}\cdot \bm{r}-2\:\!H\:\!t,
\end{align}
where $\bm{p}=m \dot{\bm{r}}$
is the canonical momentum conjugate to $\bm{r}$ and 
$H=|\:\!\bm{p}\:\!|^2/(2\:\!m)+V(\bm{r})$ is the Hamiltonian.

\end{document}